\input harvmac.tex

\def\cst {{\rm const.}}

\def \ov {\over}

\def \lr {\lref}

\def \dd {\partial}
\def \bd{\bar \bd}

\def \n {\noindent}

\def \np{ Nucl. Phys.}
\def \pl { Phys. Lett.}
\def \mpl { Mod. Phys. Lett.}

\def \pr { Phys. Rev. }
\def \cqg { Class. Quant. Grav. }
\def \jmp { J. Math. Phys.}

\lr \cghs {C. Callan, S. Giddings, J. Harvey and A. Strominger, \pr
           D45 (1992), R1005. }
\lr \garf {D. Garfinkle, G.T. Horowitz and A. Strominger, \pr D43 (1991), 
3140}
\lr \russo {J.G. Russo, L. Susskind and L. Thorlacius, \pr D46 (1992), 3444;
            \pr D47 (1993), 533.}
\lr \horowitz{G.T. Horowitz, {\it The dark side of string theory:
black holes and black strings}, 
Proceedings of the School ``String
theory and quantum gravity'', Trieste 1992, hep-th/9210119.}
\lr \fabo  {A. Fabbri and J.G. Russo, \pr D53 (1996), 6995. }
\lr \fabu {R. Balbinot and A. Fabbri, \cqg 13 (1996), 2457.}
\lr \fabd  {R. Balbinot and A. Fabbri, \cqg 14 (1997), 463.}
\lr \peleg {Y. Peleg, \mpl\  A9 (1994), 3137. }
\lr \dowqn {F. Dowker, J.P. Gauntlett, D.A. Kastor and J. Traschen, \pr D49
            (1994), 2909.}
\lr \dowci {F. Dowker, J.P. Gauntlett, S.B. Giddings and
G.T. Horowitz, \pr D50
             (1994), 2662.}  
\lr \ernst {F.J. Ernst, \jmp\  17 (1976), 515.}
\lr \gibma {G.W.Gibbson and K.Maeda, \np\  B298 (1988), 741.}
\lr \muk   {V.P. Frolov, M.A. Markov and V.F. Mukhanov, \pl\ B216 (1989), 
272.} \lr \gidstro{S.B. Giddings, {\it Quantum Mechanics of Black Holes},
Summer School in High Energy Physics and Cosmology, Trieste (1994),
hep-th/9412138; A. Strominger, {\it Les Houches Lectures on Black Holes},
Les Houches Summer School (1994),
hep-th/9501071.}
\lr \harstro{J.A. Harvey and A. Strominger,
{\it Quantum aspects of black holes}, Lectures given at the Schools ``String
theory and quantum gravity'', Trieste 1992, and TASI 92, Boulder Colorado,
 hep-th/9209055.}
\lr \bilcal {A. Bilal and C. Callan, \np\  B394 (1993) 73;
 S. de Alwis, \pl\  B289 (1992) 278; B300 (1993) 330.}
\lr \kalya{S. Kalyana Rama and S. Ghosh, \pl\ B383 (1996), 31.}

 \baselineskip8pt \Title{\vbox
{\baselineskip 6pt
\hbox{GCR-97/07/02 }
{\hbox{
   }}} }
{\vbox{\centerline {Dilaton gravity black holes with regular interior}
}}
\bigskip\bigskip\bigskip
\vskip -20 true pt
\centerline { R. Balbinot,
  L. Mazzacurati}
 \smallskip \bigskip
\centerline {\it Dipartimento di Fisica dell'Universit\`a di Bologna
and INFN sezione di Bologna}
\smallskip
\centerline {\it Via Irnerio 46, 40126 Bologna, Italy}
\smallskip
\centerline {\it emails:  balbinot@bologna.infn.it,  
mazzacurati@bologna.infn.it} 
\bigskip\bigskip\bigskip
\vskip -20 truept
\centerline {and A. Fabbri}
\smallskip\bigskip
\centerline {\it Laboratoire `Gravitation et Cosmologie
Relativistes', Universit\'e Paris VI, CNRS/URA 769}
 \smallskip
\centerline {\it Tour 22/12 4\`eme etage - Boite Courrier, 142 -
4, Place Jussieu- 75252 Paris Cedex 05, France}
\smallskip
\centerline {\it email: fabbri@ccr.jussieu.fr}
\bigskip\bigskip\bigskip
\bigskip\bigskip\bigskip
\centerline {\bf Abstract}
\bigskip
In a one parameter family of dilaton gravity theories which allow the
coupling of the dilaton to gravity and to a $U(1)$ gauge field to differ,
we have found the existence of everywhere regular spacetimes describing 
black holes hiding expanding universes inside their horizon.
\medskip
\baselineskip8pt
\noindent

\Date {July 1997}

\noblackbox
\baselineskip 14pt plus 2pt minus 2pt

\vfill \eject

\newsec{Introduction}
Two-dimensional dilaton gravity theories have become very popular toy 
models
for investigating issues related to black hole evaporation and quantum 
gravity in general (\harstro, \gidstro).
One of the most discussed is the Callan-Giddings-Harvey-Strominger (CGHS)
model \cghs\ defined by the two-dimensional action 
\eqn\strindue{
S_1^{(2)}=\int d^2 x\sqrt{-g^{(2)}}e^{-2\phi}[R^{(2)}+ 4(\nabla\phi)^2
+ 4\lambda^2 ].
}
This action is supposed to describe the s-wave perturbations along
the infinite deep throat of an extremal four dimensional black hole,
solution of the ``string originated'' low energy action 
\eqn\stringhe{
S_1^{(4)}={1\ov{2\pi}}\int d^4 x\sqrt{-g^{(4)}} e^{-2\phi}[R^{(4)} +
4(\nabla\phi)^2 -{1\ov 2}F^2]
}
where $F_{\mu\nu}$ is a $U(1)$ gauge field. 
\par \n
These magnetically charged 
extremal black holes are essentially nonsingular. The singularity
is confined to the bottom of an infinite throat along which the
geometry is effectively two dimensional, being the radius of the 
transverse two spheres nearly constant (see for instance
\harstro\ and \horowitz). 
A standard Kaluza-Klein reduction along this infinite throat
leads from the four dimensional action $S_1^{(4)}$ to the CGHS action
$S_1^{(2)}$.
\par \n
The virtue of the CGHS model is its exact solvability. Moreover, starting
from this it was also possible to construct exactly solvable models
which include the backreaction on the geometry of quantized fields
(massless scalars) \bilcal. 
The process of black hole formation and evaporation in two dimensions
should then correspond to the scattering of low-energy quanta
off (quantum mechanically) stable extremal four dimensional 
black holes. Among these two dimensional semiclassical models, one of the
most popular is due to Russo, Susskind and Thorlacius, the so called
RST model \russo.
\par
In \fabo\ it was found a simple generalization of this latter which is
still exactly integrable. The classical part of the corresponding action 
is  
\eqn\ennedue{
S_n^{(2)}= \int d^2 x \sqrt{-g^{(2)}}[e^{-{{2\phi}\ov n}}(R^{(2)}+
{4\ov n}(\nabla \phi)^2 ) +4\lambda^2 e^{-2\phi}].
}
$n$ is a parameter which we suppose here positive definite. Note that 
for $n=1$ one obtains the CGHS action $S_1^{(2)}$.
In this paper we shall find the four dimensional ``ancestor'' of the
2D theory described by $S_n^{(2)}$ and see how the value of the parameter
$n$ deeply influences the inner structure of the black hole solutions.

\vfill \eject

\newsec{Spherically symmetric black holes}

The theory we shall consider in this paper is given by the following
one parameter 4D action 
   \eqn\nquat{
S_n^{(4)}= {1\ov {2\pi}} \int d^4 x \sqrt{-g^{(4)}}
\{ e^{-2\phi}[R^{(4)}+ (6-{2\ov{n^2}})(\nabla\phi)^2] -{1\ov 2}
e^{-{{2\phi}\ov n}}F^2\} ,
}
which allows a different coupling of the dilaton with gravity and the
$U(1)$ gauge field.
This action is related to $S_1^{(4)}$ by a rescaling of the dilaton
$\phi\to {{\phi}\ov n}$ and by a conformal transformation 
$g_{\mu\nu} \to e^{2\alpha\phi} g_{\mu\nu}$ with $\alpha= 1-{1\ov n}$.
The equations of motion which follow from the action $S_n^{(4)}$ are
\eqn\motou{
R^{(4)}_{\mu\nu}-{1\ov 2}R^{(4)}g_{\mu\nu} + 2(1-{1\ov n^2})
\nabla_{\mu}\phi\nabla_{\nu}\phi +
2\nabla_{\mu}\nabla_{\nu}\phi -e^{2({{n-1}\ov n})\phi}F_{\mu\alpha}
F_{\nu}^{\alpha}+
}
$$+ g_{\mu\nu}[(1+{1\ov{n^2}})(\nabla\phi)^2 -
2\nabla^2\phi + {1\ov 4}e^{2({{n-1}\ov n})\phi)}F^2]=0,$$
\eqn\motod{
R^{(4)}-(6-{2\ov{n^2}})(\nabla\phi)^2 + (6-{2\ov {n^2}})
\nabla^2 \phi - {1\ov {2n}}F^2 e^{2({{n-1}\ov n})\phi}=0,
}
\eqn\motoc{
\nabla_{\mu}(e^{-{{2\phi}\ov n}}F^{\mu\nu})=0,
}
which are obtained by varying $S_n^{(4)}$ with respect to the metric, 
the dilaton and the Maxwell field respectively. 
\par 
Spherically symmetric magnetically charged black hole solutions 
of these equations take the form 
\eqn\busta{
ds^2= e^{2\phi_0}(1-r_0 /r)^{1-n} [-{ { (1-r_H/r)}\ov
{(1-r_0/r)}}dt^2 + {{dr^2}\ov{(1-r_H/r)(1-r_0/r)}}+r^2 d\Omega^2 ],
}
\eqn\budi{
e^{-{{2\phi}\ov n}}=e^{-{{2\phi_0}\ov n}}(1-r_0/r),
}
\eqn\maca{ 
F_{\theta\varphi}= Q\sin\theta.
}
Here $\phi_0$ is an arbitrary constant corresponding to the value of the
dilaton at infinity, 
\eqn\orri{
r_H=2M
}
is the location of the event horizon and
\eqn\singo{
r_0={{Q^2}\ov{2M}}e^{-{{2\phi_0}\ov n}}.
}
The surface $r=r_0$ represents, for $n=1$, a singularity; for other values
of $n$ we postpone relevant discussion to section III. 
\par
$Q$ represents the magnetic charge and $M$ is related to the gravitational
mass of the hole $m_g=(r_H - nr_0)/2$. They are chosen so 
that $r_0 \le r_H$ and $nr_0 < r_H$ (this last inequality ensures
the gravitational mass to be positive).
Finally, $d\Omega^2=d\theta^2 + \sin^2\theta d\varphi^2$ is the metric on
the unit two-sphere.
\par 
Extremal black holes are defined by $r_0=r_H$. In the case $n=1$ 
these are well known \horowitz\ to be completely regular, the singularity
being pushed off to infinite proper distance. The geometry along this
``throat'' is effectively two-dimensional: the radius of the 
transverse two-sphere is approximatively constant $r\simeq r_0 = 2M$.
\par \n
We shall consider the general case ($n$ arbitrary), but solutions close
to extremality.
In view of the conformal relation between the actions $S_n^{(4)}$ and
$S_1^{(4)}$ and, therefore, between the corresponding solutions 
of the field equations, we parametrize the four dimensional metric as
\eqn\pame{
ds^2=g_{ab}^{(2)} dx^adx^b + \tilde R^2 d \Omega^2,
}
where $g_{ab}^{(2)}$ ($a,b=1,2$) is the radial part of the metric and
\eqn\parad{
\tilde R=exp{[({{n-1}\ov n})\phi + {{\phi_0}\ov n}]} r
}
with $r\simeq r_0$ (down the throat).
\par \n
Inserting the decomposition eq. \pame\ in the action $S_n^{(4)}$
of eq. \nquat\ and performing a standard Kaluza-Klein reduction
along the throat one obtains, after integration over the angular 
variables, 
\eqn\ndue{
S_n^{(2)}= \int d^2 x \sqrt{-g^{(2)}}[e^{-{{2\phi}\ov n}}(R^{(2)}+
{4\ov n}(\nabla \phi)^2 ) +4\lambda^2 e^{-2\phi}],
}
where we have defined
\eqn\lambde{
4\lambda^2\equiv {{2e^{-{{2\phi_0}\ov n}}}\ov{r_0^2}}-
{{Q^2e^{-{{4\phi_0}\ov n}}}\ov{r_0^4}}.
}
The action $S_n^{(2)}$ of eq. \ndue\ coincides exactly with
the action of the two-dimensional theory introduced in ref. \fabo\ (see
our eq. \ennedue).
The correspondence between the two-dimensional black hole solutions of
the action $S_n^{(2)}$ and the near extremal four dimensional 
black hole solutions of $S_n^{(4)}$ can be easily established
by introducing the expansion parameter $\epsilon$, which measures the 
deviation from extremality \peleg\
\eqn\epesp{
\epsilon = {{r_H}\ov{r_0}}-1.
}
It is also useful to define a new spatial coordinate $\sigma$ by
\eqn\nuoco{
e^{2\lambda \sigma}=2\lambda  r.
}
In the region $r-r_0\ll r_0$, the solutions eqs. \busta, \budi\ 
can be well approximated, in the limit $\epsilon\ll 1$, by
 \eqn\apfuso{
ds^2=e^{2(1-n)\lambda\sigma}[-(1 - \epsilon
e^{-2\lambda\sigma})dt^2 + {{d\sigma^2}\ov{(1- \epsilon
e^{-2\lambda\sigma})}} + {1\ov{4\lambda^2}} d\Omega^2 ],
}
\eqn\diapp{
\phi = -n\lambda\sigma
}
where we have set, as in ref. \peleg, $r_0\equiv {1\ov{2\lambda}}$
for convenience.
\par \n
Provided one identifies the parameter $\epsilon$ with the two-dimensional
ADM mass $M$ via the relation
\eqn\idmaep{ 
\epsilon = {{M}\ov{\lambda}}
}
the radial part of the metric eq. \apfuso\ coincides with the black hole
solutions of the action $S_n^{(2)}$ (see ref. \fabo).
\par \n
Furthemore, as shown in Appendix A, also the axially symmetric
dilaton Ernst like solutions, describing quasi extremal accelerating
black holes, become, in the point particle limit, almost spherically
symmetric as the horizon is approached, of the form given by eqs.
\busta, \budi\ and \maca. On the basis of the previous discussion 
we can therefore conclude that the two-dimensional action
$S_n^{(2)}$ describes the ``low energy excitations'' along the throat
of extremal black holes, which can be either exactly spherically
symmetric or (in the point particle limit) even accelerated by an 
external field.

\newsec{Spacetime structure}

Let us now analyse the spacetime structure of the spherically
symmetric black hole solutions of $S_n^{(4)}$ given 
in the previous section.
 These solutions are labelled by the parameter $n$. This
dependence, as we shall see, introduces profound qualitative changes
in the resulting picture of the spacetime with respect to the case $n=1$
which are worth to be emphasized. 
\par \n
We rewrite for convenience the line 
element of eq. \busta\  in the form
\eqn\fumen{
ds^2=e^{2\phi_0}\big( {r\ov{r+r_0}}\big) ^{1-n}
[-(1-{{\epsilon r_0}\ov r})dt^2 + 
{{({r\ov{r_0}} +1)^2 dr^2}\ov{(1-{{\epsilon r_0}\ov r} ) 
({r\ov{r_0}})^2}} + (r+r_0)^2 d\Omega^2 ],
}
where we have simply shifted $r$ by a constant quantity $r_0$ and $\epsilon$
is given as in eq. \epesp.
The dilaton, on the other hand, reads
\eqn\fudin{
e^{-{{2\phi}\ov n}}=e^{-{{2\phi_0}\ov n}}({r\ov{r+r_0}}).
}
The range of $r$ is now  $0<r<\infty$ and the event horizon is located
at $r=\epsilon r_0$. 
\par \n
The $n$ dependence of this family of solutions is essentially contained 
in the conformal factor $({r\ov{r+r_0}})^{1-n}$, which becomes trivial
when $n=1$.
\par \n 
One should furthemore note that at infinity ($r\to\infty$),
irrespectively of the value of $n$, the metric becomes asymptotically
minkowskian and the dilaton approaches a constant value $\phi_0$.
\foot{By the way, in the axisymmetric case this is no longer true.
The metric considered in appendix A at infinity becomes conformal to the 
dilaton Melvin universe \dowci.}
\par \n 
The Ricci scalar for the metric 
eq. \fumen\ is
\eqn\scalwq{
R^{(4)}=e^{-2\phi_0}r^{n-2}(r+r_0)^{-n-3}[rr_0^2 ({1\ov 2}+
3\epsilon n + 3n - {3\ov 2}n^2) +{{r_0^3 \epsilon}\ov 2} (3n^2 -1)].
}
\par
We shall first consider extremal black holes, namely $\epsilon=0$,
for which the horizon disappears and the Killing vector 
${{\dd}\ov{\dd t}}$ is everywhere timelike.
\par \n
 In the limit $r\to 0$
from eq. \scalwq\ we obtain that $R^{(4)}\sim r^{n-1}$.
\par \n
We can then infer that for $n\ge 1$ the surface $r=0$ is
regular and the all spacetime as well.\foot{This, like all
the other results here presented, are confirmed by a careful
analysis of the other curvature invariants 
${}^{(4)}R_{\alpha\beta}{}^{(4)}R^{\alpha\beta}$ and 
${}^{(4)}R_{\alpha\beta\gamma\delta}{}^{(4)}
R^{\alpha\beta\gamma\delta}$.}
One can show that this surface lies at infinite proper distance
and for $n>1$ timelike radial geodesics never go close to it as they 
bounce back to infinity ($r\to\infty$) at turning points.\foot{Test 
particles are supposed to be free falling along the geodesics    
of the metric \fumen.}
Note also that for $n=1$ ${}^{(4)}R\to \cst$, whereas for $n>1$
${}^{(4)}R\to 0$. The difference between the two cases
is well represented in Figs. I. For $n=1$, see Fig. Ia, we have
the usual bottomless hole with constant transverse radius
($r\simeq r_0$). 
For $n>1$, on the other hand, the area of the two-sphere,
being proportional to $r^{1-n}$, increases indefinitely as $r\to 0$
(see Fig. Ib).
Note however that the gravitational mass $m_g$ measured by an asymptotic
observer at large r is negative semidefinite for $n\ge 1$.
\par \n
Finally, for $n<1$ $r=0$ is a singular surface. The singularity is 
lightlike and lies at a finite geodesic distance. It is caused by
the two spheres crushing to zero area as $r\to 0$ (see Fig. II).
\par
Let us consider now non extremal ($\epsilon\neq 0$) black holes.
These have a regular event horizon located at $r=r_H=\epsilon r_0$
which separates the space-time in an exterior (asymptotically flat)
region $r>r_H$ where the Killing vector ${{\dd}\ov{\dd t}}$
is timelike and an interior region  $r<r_H$ where 
${{\dd}\ov{\dd t}}$ is spacelike. Note that within this 
``Kantowski-Sacks'' region $r$ is a timelike variable.
\par \n
From eq. \scalwq\ the scalar curvature behaves in the limit
$r\to 0$ as ${}^{(4)}R\to r^{n-2}$. 
\par \n
For $n<2$, $r=0$ is a spacelike singularity located at finite
geodesic distance. The causal structure of the space-time
is the same as for the familiar Schwarzschild solution.
Note however that the three-dimensional spatial curvature
${}^{3}R$ of the surfaces $r=\cst$ behaves as ${}^{3}R\to r^{n-1}$.
Therefore for $n<1$ we have the usual cigar like singularity:
the area of the two spheres crushes to zero (like in Fig. II).
For $1\le n<2$ the spatial sections have regular curvature,
but the spacetime is still singular for $r=0$, the divergence
in ${}^{(4)}R$ coming from the extrinsic curvature. 
\par \n
Finally, for $n\ge 2$ the inner region is everywhere regular
and infinitely wide (like in Fig. Ib).
A static observer does not notice anything peculiar about these black 
holes: they are characterized by the mass and charge and
as usual their attractive gravitational force increases 
the more the observer is close to the horizon.
For a courageous observer willing to investigate the inner structure
of these bodies the picture appears more exciting. The 
event horizon represents as usual the no return surface, but as
he crosses it he finds himself immersed in a new expanding universe
which does not look like a Friedmann-Robertson-Walker universe:
it is still homogeneous but not isotropic.
Black holes of this kind represent wormholes to other universes, the 
door being as small as the horizon area. 
 
\newsec{Conclusions}

In this paper we have put on physical ground the 2D action $S_n^{(2)}$
introduced in ref. \fabo. This action describes perturbations
along the throat of extreme 4D dilaton black holes arising as solutions
of the action $S_n^{(4)}$.
The presence of the parameter $n$ in this latter allows the coupling 
of the dilaton to gravity and to the Maxwell field to differ.
\par 
For a range of values of $n$ we have found solutions describing
everywhere regular asymptotically flat spacetimes. The black
holes in this case hide inside their horizons infinite expanding
homogeneous universes of the Kantowski-Sacks type. 
\par \n
Caution is however required to take this conclusion as granted.
The inspection of eq.  \fudin\ reveals that the dilaton diverges
to $+\infty$ as $r\to 0$. So this region corresponds to strong
coupling and quantum effects are likely expected to alter, even
significantly, the nice classical picture here derived. 
\par
More general dilaton gravity models resulting in singularity free
solutions are discussed in ref. \kalya.

\bigskip \bigskip
  
\noindent $\underline {\rm Acknowledgements }$: R.B. and L.M. whish to
thank Richard Kerner for the kind hospitality at the Laboratoire
Gravitation et Cosmologie Relativistes where part of this work has been done.

\vfill \eject

\appendix {A} {Axially symmetric solutions}
  
As it has been shown in ref. \dowqn\ the action $S_1^{(4)}$ admits
axially symmetric solutions describing magnetically charged
black holes accelerated by a background magnetic field.
These are extensions including the dilaton field of the well
known Ernst solution \ernst. 
\par \n
In view of the conformal relation between the action $S_1^{(4)}$
and our action $S_n^{(4)}$, the axially symmetric Ernst like
solutions of this latter can be simply written as
\eqn\axsol{
ds^2={{e^{2\phi_0}\Lambda(x,y)^{1-n}}\ov{A^2 (x-y)^2}}     
({{F(y)}\ov{F(x)}})^{-n}\{ F(x)(G(y)dt^2 - {{dy^2}\ov{G(y)}})
+ F(y)( {{dx^2}\ov{G(x)}} + {{G(x)}\ov{\Lambda^2(x,y)}}d\varphi^2 )\} ,
}
\eqn\dilax{
e^{-{{2\phi}\ov n}}=e^{-{{2\phi_0}\ov n}}\Lambda(x,y){{F(y)}\ov{F(x)}},
}
\eqn\efixo{
A_{\varphi}=- {{e^{{{\phi_0}\ov n}}}\ov{B \Lambda (x,y)}}
[1+Bqx] + k
}
where
\eqn\spsimb{
\Lambda(x,y)=[1+Bqx]^2 + {{B^2}\ov{2A^2 (x-y)^2}}G(x)F(x),
}
\eqn\effegi{
 F(\xi)=(1+r_-A\xi), \ \ \ G(\xi)=(1-\xi^2 - r_+ A \xi^3).
}
For a detailed description of these solutions we refer to
refs. \dowqn, \dowci. 
\par
The metric contains essentially four parameters $r_+, r_-, A$ and $B$.
$r_+$ and $r_-$ are related to the ADM mass and total charge 
respectively, whereas $A$ and $B$ to the acceleration of the hole
and to the strength of the external magnetic field.
Following \dowqn\  the parameters $r_+$ and $A$ are chosen such
that $r_+A<2/(3\sqrt{3})$, so that the function $G(\xi)$
has three real distinct roots. Call these roots $\xi_2$, $\xi_3$ and 
$\xi_4$ and further define
$\xi_1\equiv - {1\ov {r_- A}}$. In the parameter region where
$\xi_1\leq\xi_2<\xi_3<\xi_4$ the surface $y=\xi_2$ represents the
black hole horizon and $y=\xi_3$ the   
acceleration horizon. 
\par \n
$x$ and $\varphi$ are angular coordinates,
$\xi_3 \le x \le \xi_4$ and $0\le \varphi \le
{{4\pi\Lambda(\xi_3)}\ov{G^{'}(\xi_3)}}$. This last restriction comes 
from the requirement that the metric has no nodal singularity, 
 i.e.
\eqn\nonosi{
G^{'}(\xi_3)\Lambda (\xi_4)= -G^{'}(\xi_4)\Lambda (\xi_3).
}
The purpose to exploit in detail the form of these solutions is
because one can show that along the throat of nearly extremal 
accelerating black holes (extremality is here defined as
$\xi_1=\xi_2$) the axially symmetric metric eq. \axsol,
the dilaton eq. \dilax\ and the Maxwell field \efixo\
approach, as $y\to\xi_2$, the spherically symmetric simple form
of eqs. \busta, \budi\ and \maca\ provided that $r_+A\ll 1$.
In the case of exact extremality this last restriction on the 
parameters is not required.
\par \n
To this end, let us introduce the quantity 
\eqn\epaxi{
\epsilon \equiv {{\xi_1 - \xi_2}\ov{\xi_2}}
} 
which measures deviation from extremality. We suppose $\epsilon\ll 1$.
One can then write from eqs. \effegi\ in the limit $y\to\xi_2$
\eqn\appfu{
F(\xi) = r_-A (\xi -\xi_2 (1+\epsilon)),
}
\eqn\monsgi{
G(\xi)=-(r_+A)(\xi - \xi_2)(\xi - \xi_3)(\xi - \xi_4).
}
\par \n
Moreover, using the regularity condition eq. \nonosi\ and 
\appfu, \monsgi\
we get
\eqn\nonod{
1+ 2Bq\xi_2 + B^2q^2 (\xi_2(\xi_3 + \xi_4)-\xi_3\xi_4)=0.
} 
Using this last result we can rewrite $\Lambda$ as
\eqn\lacor{
\Lambda(x,y)=\alpha (x-\xi_2) - B^2q^2 (x-\xi_3)(x-\xi_4) 
{{(-\xi_2\epsilon)}\ov{x-\xi_2}},
}
where we have defined
\eqn\alpa{
\alpha= 2Bq + B^2q^2 (\xi_3 + \xi_4).
}
As in ref. \dowci\ we can now define the new coordinates $r, t^{'},
\theta$ and $\tilde\varphi$
\eqn\coer{
y={{\xi_2 \hat r_+}\ov {r}},
}
\eqn\coeru{
t={{t^{'}}\ov{\sqrt{r_+r_- (-\alpha\xi_2)
(\xi_4 - \xi_2)(\xi_3 - \xi_2) } }} ,
}
\eqn\defas{
x={1\ov 2} [\xi_3 + \xi_4 - {{\xi_4 - \xi_3 +(\xi_4 +\xi_3 -2\xi_2)
\cos\theta}\ov{\cos\theta + (\xi_4 + \xi_3 -2\xi_2)/(\xi_4 - \xi_3)}}],
}
\eqn\nuang{
\varphi={{2\Lambda (\xi_3)}\ov{G^{'}(\xi_3)}}\tilde\varphi,
}
where 
\eqn\recap{
\hat r_+={1\ov A} \sqrt{ {{r_- (-\alpha \xi_2)}\ov {r_+ (\xi_4 -
\xi_2)(\xi_3 - \xi_2)}}}. 
}
At this point the metric eq. \axsol, the dilaton eq. \dilax\
and the Maxwell field \efixo\ in terms of the new variables read
\eqn\fumene{
ds^2=e^{2\phi_0}(-\xi_2\alpha)^{-n}
[1+g(x) {{\epsilon}\ov{\xi_2 (1- {x \ov{\xi_2 }})^2}}]^{1-n}
}
$$
\{ \ \{ (1+\epsilon)^{-n}(1+
{{\epsilon}\ov{1- {x\ov{\xi_2}}}})^{(1+n)}
(1-{ {\hat{r}_+} \ov r }) [1-{ {\hat{r}_+} \ov r } (1-\epsilon)]^{-n}
\}
[-dt^{'2} + { {dr^2} \ov {(1-  { {\hat{r}_+}\ov{r} }  )^2} } ]  +  $$
$$(1+\epsilon)^{1-n} (1+
{ {\epsilon}\ov{1-{x\ov{\xi_2}} }} )^{n}
(\hat{r}_+^2) [1- { {\hat{r}_+}\ov{r} }(1-\epsilon)]^{1-n}
[d\theta^2 + \sin^2 \theta d\tilde\varphi^2
(1- {{2g(x)}\ov{\alpha}} {{\epsilon}\ov{\xi_2}}  {{1} \ov {(1- 
{x\ov{\xi_2}} )^2 }} )] \  \} , $$
\eqn\difuap{
e^{-{{2\phi}\ov n}} =
e^{-{{2\phi_0}\ov n}} (-\xi_2 \alpha )
[1+{{g(x)}\ov{\alpha}}{{\epsilon}\ov{\xi_2 (1- {x\ov{\xi_2}})^2}}]
{{( 1+\epsilon)}\ov{(1+  { {\epsilon}\ov{1- { {x}\ov{\xi_2} } }} )}}
[1-{ {\hat{r}_+} \ov r } (1-\epsilon)],
}
\eqn\esemf{
A_{\tilde\varphi}=[ - {{e^{{{\phi_0}\ov n}}(1+Bqx)}\ov{B\alpha (x-\xi_2)
[1+{{g(x)}\ov{\alpha}}{{\epsilon}\ov{\xi_2 (1- {x\ov{\xi_2}})^2}}]}}
+ k]{{2\Lambda(\xi_3)}\ov{G^{'}(\xi_3)}}
}
where
 $g(x)=B^2q^2 (x-\xi_3)(x-\xi_4)$ and $x$ is considered as a function 
of $\theta$ by eq. \defas. 
\par \n
The result is straightforward in the exact extreme case ($\epsilon=0$)
as all $x$ dependent parts in eqs. \fumene-\esemf\ drop off \dowci.
In the quasi extreme case ($\epsilon\ll 1$)
let us consider the point particle limit (or equivalently the limit
of small acceleration) by defining
\eqn\smpar{
\tau=r_+A.
}
In the limit $\tau\ll 1$ it is easy to see that the $x$ dependent 
parts of the metric eq. \fumene\ are of order $O(\tau\epsilon)$,
this because $\xi_2\sim - {1\ov{\tau}}$.
To order $O(\epsilon)$ the metric reads
\eqn\lorde{
ds^2=
e^{+2\phi_0 } (-\xi_2 \alpha )^{-n} (1+\epsilon) \{
(1-{ {\hat{r}_+} \ov r }) [1-{ {\hat{r}_+} \ov r } (1-\epsilon)]^{-n}
}
$$
[-dt^{'2} + { {dr^2} \ov {(1-  { {\hat{r}_+}\ov{r} } )^2} } ]  +
\hat{r}_+^2 [1- { {\hat{r}_+}\ov{r} }(1-\epsilon)]^{1-n}
d\Omega^2 \}
$$
which defining $ \hat{r}_- \equiv \hat{r}_+ (1-\epsilon) $
becomes
\eqn\cacca{
ds^2=
e^{2\phi_0 } (-\xi_2 \alpha )^{-n} (1+\epsilon)
[ 1- { {\hat{r}_-} \ov {\hat{r} } }]^{1-n}
 \{
-{{(1-{ {\hat{r}_+} \ov{r} })}\ov{
( 1- { {\hat{r}_-} \ov {\hat{r} } })}}dt^{'2} +
 { {dr^2} \ov {(1-  { {\hat{r}_+}\ov{r} })( 1- { {\hat{r}_-} \ov 
{r}})}}
+  \hat{r}_+^2 d\Omega^2 \} .
}
Similarly, for the dilaton and Maxwell field we find
\eqn\diorde{
e^{-{{2\phi}\ov  n}} =
e^{-{{2\phi_0}\ov  n}} (-\xi_2 \alpha )
[1-{ {\hat{r}_-} \ov r } ],
}
\eqn\curio{
A_{\tilde\varphi}=\hat q (1-\cos\theta)
}
where $\hat q\equiv {{\hat r_+ }\ov{\sqrt{2}}}e^{{{\phi_0}\ov n}}
(-\alpha\xi_2)^{-1/2}$.
The quasi extremal ($\epsilon\ll 1$) point particle ($\tau\ll 1$)
axially symmetric solutions therefore become, as the horizon is
approached, spherically symmetric of the simple form in eqs. 
\busta, \budi\ and \maca, with the obvious replacement
of $r_0$ and $r_H$ with $\hat r_-$ and $\hat r_+$ respectively.

\vfill \eject

\listrefs

 \vfill \eject

Figure caption:
\bigskip
\bigskip \noindent
Fig. I: qualitative picture of
\par \n
a) spatial section of the $n=1$ extreme ($\epsilon=0$) black hole
\par \n
b) spatial section of the $n>1$ extreme black hole
\bigskip \bigskip \noindent
Fig. II: spatial section of the $n<1$ black hole

\vfill \eject

\end